\documentclass[twocolumn,aps,amsmath,amssymb,pra]{revtex4-1}

\usepackage{bm}
\usepackage{xcolor}
\usepackage{graphicx}
\usepackage{boxedminipage}
\allowdisplaybreaks[4]

\newcommand{\w}{\omega}

\newcommand{\la}{\langle}
\newcommand{\ra}{\rangle}

\newcommand{\be}{\begin{equation}}
\newcommand{\ee}{\end{equation}}
\newcommand{\bea}{\begin{eqnarray}}
\newcommand{\eea}{\end{eqnarray}}
\newcommand{\bsube}{\begin{subequations}}
\newcommand{\esube}{\end{subequations}}
\newcommand{\Eq}[1]{Eq.\,(\ref{#1})}

\newcommand{\Fig}[1]{Fig.\,\ref{#1}}

\newcommand{\comments}[1]{}

\allowdisplaybreaks[1]

\begin{document}
\title{On the relation between momentum uncertainty and thermal wavelength}

\author{Zi-Fan Zhu}

\author{Yao Wang} 
\email{wy2010@ustc.edu.cn}
\affiliation{University of Science and Technology of China, Hefei, Anhui 230026, China}

\date{\today}

\begin{abstract}

For quantum particles in a Boltzmann state, we derive an inequality between momentum uncertainty $\Delta p$ and thermal de Broglie wavelength $\lambda_{\rm th}$, expressed as $\Delta p \geq \sqrt{2\pi}\hbar/\lambda_{\rm th}$, as a corollary of the Boltzmann lower bound for the Heisenberg uncertainty product proposed in the previous work [EPL, {\bf 143}, 20001 (2023)].
\end{abstract}
\maketitle

\section{Introduction}

Quantum uncertainty is a fundamental aspect of quantum mechanics that highlights the intrinsic limitations in measuring certain pairs of physical properties of particles with arbitrary precision \cite{Hei49}. 
Central to this is Heisenberg's uncertainty principle, which articulates that the product of the uncertainties in position ($\Delta x$) and momentum ($\Delta p$) measurements is always greater than or equal to a specific value ($\hbar/2$), where 
$\hbar$ is the reduced Planck's constant.
This principle arises from the wave-particle duality of matter, where particles exhibit both wave-like and particle-like properties. Consequently, a particle's wave function, which describes the probabilities of its position and momentum, cannot be sharply defined in both domains simultaneously. Thus, precisely measuring a particle's position results in a large uncertainty in its momentum and vice versa. This inherent uncertainty is not due to any shortcomings in measurement technology but is a fundamental characteristic of nature itself, profoundly affecting our understanding of the microscopic world and the behavior of particles at quantum scales. Experimental evidence, such as electron diffraction and the double-slit experiment, vividly demonstrates these principles, illustrating the deep and counterintuitive nature of quantum reality \cite{Hei27172,Dir30,Fey70}.

On the other hand, the thermal de Broglie wavelength is a fundamental concept in statistical thermodynamics that serves as a crucial link between quantum mechanics and classical thermodynamics, encapsulating the characteristic quantum mechanical wavelength of a particle at a given temperature. Defined by the formula 
\be 
\lambda_{\rm th}\equiv\sqrt{\frac{2\pi\hbar^2}{mk_{B}T}},
\ee
with  $m$ the particle's mass and $k_B$ the Boltzmann's constant, and 
$T$ the absolute temperature, this wavelength reflects the scale at which quantum effects become significant \cite{Hua87,Cha87}.
At low temperatures or for particles with small masses, the thermal de Broglie wavelength is relatively large, indicating the prominence of quantum effects such as wave-particle duality and the necessity of quantum statistics (Bose-Einstein or Fermi-Dirac). Conversely, at high temperatures or for heavier particles, this wavelength is small, and classical mechanics suffices to describe the system's behavior. The thermal de Broglie wavelength is instrumental in determining the quantum versus classical behavior of gases, influencing the formulation of partition functions, and understanding phase transitions like Bose-Einstein condensation and superfluidity. This concept is vital for accurately describing the thermodynamic properties of systems across different temperature regimes and particle masses \cite{Hua87,Cha87}.

In recent years, there has been growing interest in characterizing the 
quantum state at finite temperatures 
in thermodynamic regimes \cite{Lee18032119,Bra18090601,Has19110602,Tim19090604,Sei19176,Has19062126,Gua19033021,Hor2015,Van22140602,Fu22024128,Kam23L052101}.
Regarding fluctuations, it is widely believed that finite-temperature fluctuations override quantum fluctuations.
This point can be observed via a recently proposed equality for arbitrary quantum Boltzmann state \cite{Wan2320001}, which reads
\be \label{BUR}
\Delta x \Delta p \geq \frac{\hbar}{2} \times \Gamma \Big( \frac{1}{4\pi r^2} \Big).
\ee
Here, 
$
r \equiv \Delta x / \lambda_{\rm th}
$
is the ratio of $\Delta x$ to the thermal de Broglie wavelength $\lambda_{\rm th}$, and the function $\Gamma(x)$ is constructed as
\be \label{gammax}
\Gamma(x) = g^{-1}(x)/x,
\ee
with $g^{-1}$ being the inverse function of $g(x) = x \tanh(x/2)$.
The right-hand side of \Eq{BUR} is termed the Boltzmann lower bound, differing from the Heisenberg lower bound by a factor of 
$\Gamma(z)$ with $z\equiv (4\pi r^2)^{-1}\geq 0$, which is always greater than $1$. In particular, at high temperatures, 
$\Gamma(z)\gg 1$, highlighting how finite-temperature fluctuations dominate over quantum fluctuations.
To completeness, we  outline the derivation of \Eq{BUR} in the Appendix.

\section{Inequality}
In this section, we will derive the central result of this work: an inequality between momentum uncertainty $\Delta p$ and thermal de Broglie wavelength $\lambda_{\rm th}$, expressed as \be   
\Delta p \geq \sqrt{2\pi}\hbar/\lambda_{\rm th}. 
\ee
To proceed, we recast \Eq{BUR} as 
\be 
 \Delta p \geq \frac{\hbar}{2\Delta x} \Gamma(z)=\frac{\sqrt{2\pi}\hbar}{\lambda_{\rm th}}(2 z)^{-1/2}g^{-1}(z).
\ee
\begin{figure}[h]  
\centering\includegraphics[width=\columnwidth]{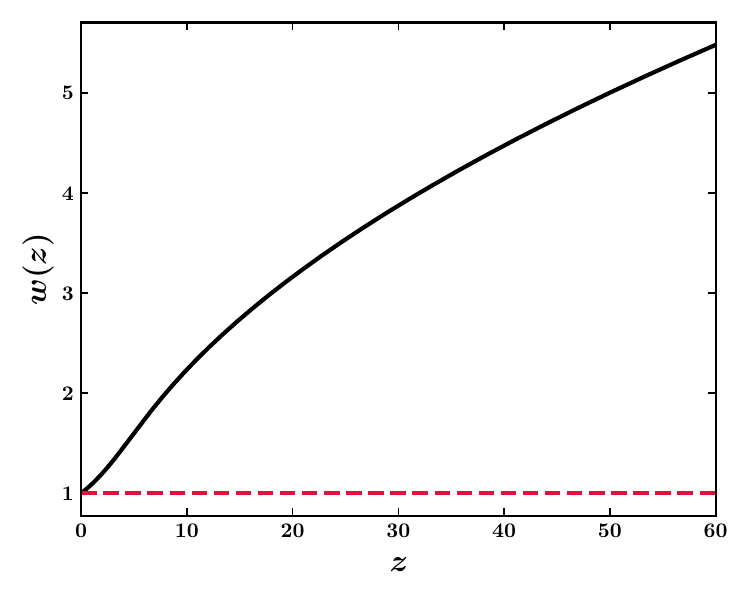} 
\caption{The image of function $w(z)$; cf.\,\Eq{tbp}.}      
\label{fig1}   
\end{figure}
This, combined with the mathematical inequality 
\be \label{tbp}
w(z)\equiv (2 z)^{-1/2}g^{-1}(z)\geq w(0)=1,
\ee
directly leads to
\be \label{key}
\Delta p \geq \frac{\sqrt{2\pi}\hbar}{\lambda_{\rm th}}\equiv \frac{p_{\rm th}}{\sqrt{2\pi}},
\ee
where the thermal momentum is defined as $p_{\rm th}\equiv 2\pi\hbar/\lambda_{\rm th}$.
We demonstrate  the  validity of \Eq{tbp} in \Fig{fig1}.
Equation (\ref{key}) is the key result of this work, which claims that for a quantum particle in Boltzmann states, the $p_{\rm th}/\sqrt{2\pi}$ serves as the lower bound for the momentum uncertainty.

\section{Summary}
To summarize, in this work we establish a relationship between the uncertainty in momentum ($\Delta p$) and the thermal de Broglie wavelength ($\lambda_{\rm th}$), expressed in the inequality $\Delta p \geq \sqrt{2\pi}\hbar/\lambda_{\rm th}$, which is consequence of the Boltzmann lower bound \cite{Wan2320001}.
This inequality provides a fundamental insight into the interplay between quantum uncertainty and thermal effects. It highlights how the thermal de Broglie wavelength influences the lower bound of momentum uncertainty, emphasizing the significance of temperature in quantum mechanical measurements\cite{Lan54643,Elc57161,Koc81380,Mig202471,Koy22014104,Gar04,Cle101155,Soa14825}.
It could also help understandings thermal  behaviors of complex systems  in a wide array of disciplines. \cite{Lee071462,Che09241,Lam1310}. 
Future research can explore practical implications in quantum systems and advance our understanding of how thermal fluctuations impact quantum phenomena, potentially leading to novel applications in quantum technologies and precision measurements.

Moreover, analyses based on thermodynamic considerations have shown that the Heisenberg uncertainty relation undergoes deformation in the quantum gravity regime \cite{Buo22136818}.
The same techniques employed in this study can be similarly applied there, help elucidating the gravity-thermodynamic conjecture \cite{Jac951260}.

\begin{acknowledgments}
  Support from
the National Natural Science Foundation of China (Grant Nos.\ 22103073 and 22373091) and the USTC New Liberal Arts
Fund (No.\ FSSF-A-230110) is gratefully acknowledged.    
\end{acknowledgments}

 \appendix 
 \section{Derivation of \Eq{BUR}}
In this appendix, we derive the \Eq{BUR}, following the Ref.~\onlinecite{Wan2320001}.
 The methodology 
closely related to that exploited in deriving thermodynamic uncertainty relations \cite{Str22}.
Denote
\be \label{ct}
c(t)\equiv \la \delta \hat x(t)\delta \hat x(0)\ra,
\ee
where the average $\la(\,\cdot \,)\ra$ runs over the Boltzmann state $\propto\exp(-\beta \hat H)$. 
Evidently, we have
\be\label{c0} 
c(t=0)=(\Delta x)^2.
\ee
If the kinetic energy is term as $\hat p^2/(2m)$, we further have 
\be  \label{dc0}
 \dot c(0)=\frac{1}{m}\la \delta \hat p\delta\hat  x\ra=-\frac{1}{m}\la \delta \hat x\delta \hat p\ra=-\frac{i\hbar} {2m},
\ee 
and 
  \be \label{ddc0}
  \ddot c(0)=-\bigg(\frac{\Delta p}{m}\bigg)^2.
  \ee

The derivation of \Eq{BUR} starts with the detailed balance relation reading \cite{Cal5134, Yan05187}
\be \label{db}
  \frac{C(-\w)}{ C(\w)}=e^{-\beta \hbar \w},
\ee
where $
C(\w)\equiv\frac{1}{2}\int_{-\infty}^{\infty}\!{\rm d}t\,e^{i\w t} c (t)
$
is  real and positive definite.
Therefore, one may introduce a probability distribution $P(\w)$ over the frequency domain as
\be \label{P}
P(\w)\equiv \frac{C(\w)}{\pi (\Delta x)^2},\ \ \ \ \w\in(-\infty,\infty).
\ee
The normalization factor $\pi (\Delta x)^2$ can be obtained from
the inverse transform  
$
c(t)=\frac{1}{\pi}\int_{-\infty}^{\infty}\!{\rm d}\w\,e^{-i\w t}C(\w),
$
by setting $t=0$.
Equation (\ref{db}) then directly leads to 
\be \label{db4}
P(-\w)=e^{-\beta \hbar \w}P(\w).
\ee

Now we further introduce a  distribution  defined on the interval $\w\in[0,\infty)$ as \cite{Str22}
\be 
Q(\w)=(1+e^{-\beta \hbar \w})P(\w).
\ee
By using \Eq{db4}, it is easy to verify $Q(\w)$ is also a normalized distribution in $\w\in[0,\infty)$.
One may denote expectation values of any
function $f(\w\geq 0)$ with respect to $P(\w)$ by $\la f(\w) \ra$, while  denote that to $Q(\w)$ by $\la f(\w) \ra_{\rm Q}$. 

The first and second moments of $\w$ with respect to $P(\w)$ and $Q(\w)$ are then connected as
\be \label{1stm}
\beta \hbar \la \w\ra=\la g(\beta \hbar \w)\ra_{\rm Q}
\ee
and
\be \label{2ndm}
 \la \w^2 \ra=\la \w^2 \ra_{\rm Q}.
\ee
with 
$g(x)$ as defined in \Eq{gammax}.

The key step is to verify $g(\sqrt{x})$ is be  monotonically increasing and concave for $x\geq 0$ \cite{Str22}, which leads to that $k^2$, with $k=g^{-1}$, must be convex.
Then according to Jensen's inequality \cite{Jen06175}, we have
\be\label{key}
\begin{split}
\la \w^2 \ra&\overset{(\ref{2ndm})}{=\joinrel=}\la \w^2 \ra_{\rm Q}=\frac{\la k^2[g(\beta \hbar \w)] \ra_{\rm Q}}{(\beta \hbar)^{2}}
\\ &
\geq \frac{k^2[\la g(\beta \hbar \w) \ra_{\rm Q}]}{(\beta \hbar)^{2}}
\overset{(\ref{1stm})}{=\joinrel=}\frac{k^2(\beta  \hbar  \la \w\ra)}{(\beta \hbar)^2}.
\end{split}
\ee
The second equality in the first line is deduced from $[k g(\sqrt{x})]^2=x$, by setting $x=(\beta \hbar \w)^2$.
The ``$\geq$'' in the second line is the Jensen inequality for the convex function $k^2$.

Equations (\ref{ct})--(\ref{P}) give  rise to
\be 
\la  \w\ra=\frac{\hbar}{2m(\Delta x)^2}
\ee
and
\be 
\la  \w^2\ra=\Big(\frac{\Delta p}{m\Delta x}\Big)^2.
\ee
Then after some simple algebra, \Eq{key} gives \Eq{BUR}, with $ 
\Gamma(x)$ as defined in \Eq{gammax}.
It is easy to see $\Gamma(x)> 1$, since $\tanh(x/2)< 1$ for all $x>0$.
 %
 %\bibliographystyle{./aiptit}
 %\bibliography{./bibrefs}

\end{document}